# Optically active beams: non-reciprocal optical activity in free space induced by spin-orbital interaction of light

Sheng Liu[1,*], Shuxia Qi,[1] Peng Li,[1] Bingyan Wei,[1] Peng Chen,[2] Wei Hu,[2] Yi Zhang,[1] Xuetao Gan,[1] Peng Zhang,[3] Yanqing Lu,[2] Zhigang Chen,[4] and Jianlin Zhao[1,*]

[1]MOE Key Laboratory of Material Physics and Chemistry under Extraordinary Conditions, and Shaanxi Key Laboratory of Optical Information Technology, School of Science, Northwestern Polytechnical University, Xi'an 710129, China
[2]National Laboratory of Solid State Microstructures, Key Laboratory of Intelligent Optical Sensing and Manipulation, Collaborative Innovation Center of Advanced Microstructures, and College of Engineering and Applied Sciences, Nanjing University, Nanjing 210093, China
[3]State Key Laboratory of Transient Optics and Photonics, Xi'an Institute of Optics and Precision Mechanics, Chinese Academy of Sciences, Xi'an 710119, China
[4]The MOE Key Laboratory of Weak-Light Nonlinear Photonics, TEDA Applied Physics Institute and School of Physics, Nankai University, Tianjin 300457, China

*shengliu@nwpu.edu.cn; jlzhao@nwpu.edu.cn*

**Optical activity, the power of a medium to rotate the polarization of a light beam, has contributed significantly to molecular structure assessments in stereochemistry, biomolecular science and crystallography. Thus far, it is commonly believed that optical activity is manifested only in the chiral media which can give rise to circular birefringence of light. Here, we experimentally demonstrate that free space can also support the implementation of non-reciprocal optical activity with Bessel beams by spin-orbital interaction. Specifically, non-diffractive optically active beams are realized, with their optical rotatory power readily controlled by simple optical elements. We show that such free-space optical activity can be exploited to form non-reciprocal optical components such as polarization rotators, isolators, and circulators. Our results may bring about new possibilities of media-independent optical activity to other transverse waves ranging from radio to optical frequencies.**

# INTRODUCTION

Optical activity (OA), referred to rotatory polarization, is the ability to rotate the polarization of light along the propagation direction. It was discovered at the beginning of the 19th century and is now being extensively studied due to its important applications in analytical chemistry, biology, and crystallography. Traditionally, it was considered that OA occurs in the media including chiral molecules, such as sucrose, quartz, and some protein molecules with helical structure. The chiral structures of molecular units, associated with the mirror asymmetry, are considered crucial in exciting OA. In recent years, several types of media with strong chirality have been artificially constructed with metamaterials, and offer broadband and nondispersive characteristics, or gigantic polarization rotatory power[1-5]. Yet it is demonstrated that the non-chiral metamaterials can also perform the effect of OA, by introducing extrinsic chirality with oblique incidence[6,7], or by manipulating the phase difference between the two circularly polarized components[8]. In principle, OA is considered a product of circular birefringence, relying much on the optically active materials. Apart from the OA in bulk media and thin metamaterials, the circular birefringence can be artificially manifested from the topological Berry phase in helical coiled monomode optical fibers [9-11], according to the spin-orbit interaction of photons during trajectory variation[12]. With this revelation, the OA is effectively induced in optical fibers with other twisted structures likewise [13-15].

    The polarization rotation is irrelevant to the incident direction of light in the chiral media, or even metamaterials with chiral structures[1,3]. Such reciprocity owes to the linear field-matter interaction permeated in physics wildly. To break the

reciprocity, magneto-optical materials subjected to magnetic fields are generally used. Such non-reciprocal OA is known as Faraday effect[16], and can be enhanced in magneto-optical waveguides[17,18] and magneto-plasmonic photonic crystal[19]. It provides the basis of non-reciprocal components such as isolators, gyrators, and circulators, and enable the applications in wireless communication and all-optical information processing [20–23].

Until now, the reported OAs are realized by light-matter interaction in specific materials or setups, which suffer limitations in some applied fields. Hence, lifting the restriction of media on OA is significant to expend the application range. Thus far, there have been no reports that free space can support the circular birefringence to generate OA, to the best of our knowledge. Here we propose a solution with spin-orbital interaction based on the theories of Gouy phase shift and Pancharatnam-Berry phase. A specific liquid crystal (LC) conical-wave plate (CWP) is designed to realize the circular birefringence of Bessel beam in free space. After passing through the CWP, a linearly polarized Bessel beam exhibits non-reciprocity polarization rotation as it travels in free space (see Fig. 1). Namely, it is transformed into a beam with OA which possesses power to rotate the polarization plane by itself. The optical rotatory power is dependent on the radial wave vector $k_\rho$ of Bessel beam and the radial frequency $\delta_k$ of the CWP, and can be enhanced or declined via specific optical elements. Moreover, we demonstrate the response of the OA to an index change, and the capability of forming non-reciprocal optical components such as polarization rotator, isolator, and circulator.

## Results

**Circular birefringence in free space**

To realize the OA in free space, we first achieved the phenomenon akin to circular birefringence which offers a phase difference of spin states of light (circular polarizations) during propagation. As a matter of fact, the variation in free-space propagations of different spin states is manifested in the isotropic inhomogeneous medium or structures by spin-orbital interaction [24,25] as the form of, for example, spin Hall effect [26–28], spin-controlled shaping [29,30], and spin-directional coupling [31–33]. One effective technique to realize and even enhance the spin-orbital interaction of light is exploiting the Pancharatnam-Berry phase arising in polarization manipulations [34–36]. An enhanced spin-orbital interaction enables a great variation between the wave vectors of the two spin states, which guides the modulation of the orbital angular momentum.

Here, we introduce a specially designed Pancharatnam-Berry phase element, LC conical-wave plate (CWP, see Methods), to generate a spin-dependent bifurcation of phase velocity decided by the longitudinal wave vector. This element can be considered as a half wave plate with its local fast axes arranged conically. Namely, the orientation angle meets $\Theta=-\delta_k\rho/2$, where $\delta_k$ is a constant describing the frequency of the axis orientation change along radial coordinate $\rho$. The CWP has different responses for the two photonic spin states: it reverses the spins and attaches different Pancharatnam-Berry phases to them. For instance, the right- and left-handed spin states $|\sigma_+\rangle$ and $|\sigma_-\rangle$ are transformed into left- and right-handed ones $|\sigma_-\rangle$ and $|\sigma_+\rangle$, with phases $\exp(\pm i\delta_k\rho)$ appended, respectively (see Methods). A light wave with a Bessel profile $J_0(k_\rho\rho)$ or with a conical phase $\exp(ik_\rho\rho)$ can be considered as a superposition of a series of plane waves with wave vectors located on a cone with radius $k_\rho$. Thus a circularly polarized Bessel beam passing through the CWP performs wave vector change as shown in Fig. 1b. The wave vector cones for left- and right-handed spins expand and contract respectively, with their slant heights keeping a constant $k_0 = 2\pi/\lambda$. Correspondingly, the wave vectors along propagation direction $k_z$, which directly decides the phase velocity of Bessel beam, are decreased and increased, presenting a birefringence for the two circular polarizations.

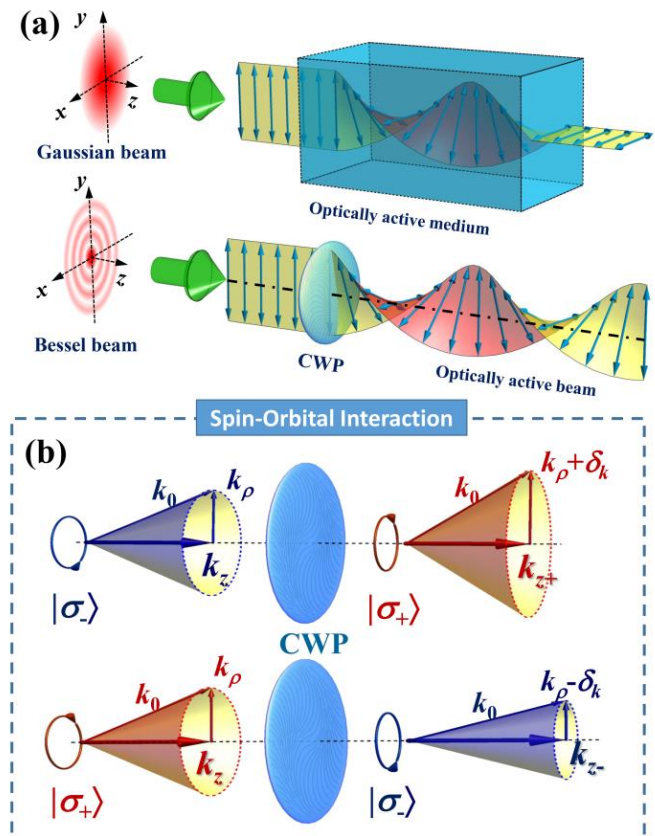

**Figure 1 Schematic illustration of optical activity (OA) in free space.** (a) Sketch of the traditional (top) and free-space OA (bottom), where the blue arrowheads denote the polarization orientations. A linelly polarized beam performs polarization rotation in an active medium own to the circular birefringence. Similarly, the polarization rotation of Bessel beam occurs in free space after passing through a conical-wave plate (CWP). (b) Spin-orbital interaction of Bessel beam triggered by CWP. The spin states passing through the CWP change their sign, leading to the conversion of wave vector. As a result, the velocity which depends on the axial wave vector suffers a spin-related bifurcation, i.e. circular birefringence.

The velocity difference of circular polarizations can be quantificationally analyzed by the Gouy phase shift, which is a common effect of laser beam denoting the additional phase shift of a light beam along propagation direction with respect to a plane wave[37]. It is notable that the Gouy phase shift of Bessel beam linearly varies with the propagation distance[38], i.e. $\varphi_{Gouy}(k_\rho)=k_\rho^2 z/(k_0+\sqrt{k_0^2-k_\rho^2})$. Thus right- and left-handed polarized Bessel beams passing through the CWP acquires a phase difference $\Delta\varphi=\varphi_{Gouy}(k_+)-\varphi_{Gouy}(k_-)$, where $k_\pm=k_\rho\pm\delta_k$.

**Optical activity in free space**

For a monochrome input zero-order Bessel beam with linear polarization propagating along $z$-axis, the complex amplitude is described by $\mathbf{E}_{in}=J_0(k_\rho\rho)|\mathbf{L}(\theta_0)\rangle$, where $|\mathbf{L}(\theta_0)\rangle=[\cos\theta_0, \sin\theta_0]^T$ denotes the linear polarization along angle $\theta_0$. When it is input to the CWP, the resulting field presents polarization rotation due to the circular birefringence as schematically shown in Figure 1c, as approximately expressed as (see Methods)

$$|\mathbf{E}_{out}\rangle = J_0\left(k_\rho\rho\right)\left|\mathbf{L}\left(\frac{k_\rho\delta_k}{k_0}z-\theta_0\right)\right\rangle. \qquad (1)$$

This equation indicates that a propagate-rotating polarization is introduced to the linearly polarized Bessel beam by the CWP, and it has striking similarities with the traditional OA: (1) Linear rotation: the rotation angle of polarization direction is proportional to the propagation distance $z$. Accordingly, we defined the specific rotation $[\alpha]$ here as the rotation angle in unit propagation length, i.e. $[\alpha]=k_\rho\delta_k/k_0$. (2) Circular birefringence: the specific rotation is proportional to the radial frequency of the CWP $\delta_k$, which can be viewed as a description of the amount of circular birefringence in the optically active medium. (3) Optical rotatory dispersion: the specific rotation is proportional to the wave vector $k_\rho$ of the Bessel beam which decides the phase velocity, representing the response difference of CWP for $k_\rho$. (4) Non-reciprocity (also see Figure 2): different from the optically active medium, the injections from the front and back sides of CWP have a sign reversal of $\delta_k$, resulting in the dextrorotation and levorotation (clockwise and counterclockwise polarization rotation from the view of the observer).

**Experiment**

The experiments for realizing OA in free space are performed with the optical setup shown in Figure 2a. A laser beam (He-Ne laser, 632.8 nm) is collimated by the reversed telescope (RT), with its polarization state adjusted to be linear by a polarizer $P_1$. The beam is transformed into a Bessel mode via an axicon (Ax), and then passes through the CWP ($\delta_k=\pi$ mm$^{-1}$) fabricated in a typical birefringent material LC (see Methods). To monitor the propagating process of the light, the intensity distribution of the output beam is recorded by a CCD camera moved along the propagation direction step by step. The polarizer $P_2$ is used to measure the rotation angle of polarization. The propagation process of the beam through the axicon and CWP within a distance of 16cm, which is displayed by the three-dimensional intensity distribution of the recorded beam profiles in Figure 2b, exhibits a non-diffraction optical beam with Bessel profile. When this propagation process is analyzed by the analyzer $P_2$, the tailored Bessel beam performs a periodic variation in intensity. Figure 2c shows the evolution processes of light intensities after passing the horizontal (top) and vertical (bottom) analyzers, which describe the horizontally ($E_x$) and vertically ($E_y$) polarized components of the light beam, respectively. The staggered periodic extinction in these two pictures reveals the polarization rotation of light. In order to measure the rotation angle $\theta$ of the polarization, we rotate the analyzer P$_2$ to detect the angle for extinction in different propagation distances. The rotation angle versus propagation distance is shown in Figure 2e (levorotation), which is changing in a perfectly linear fashion. The pink solid line in Figure 2e shows the data fitting, where the slope (i.e. the specific rotation [$\alpha$]) is measured 1.117°/mm, coinciding well with the theoretical prediction.

**Non-reciprocity**

The polarization rotation orientation in an active medium depends on the chirality of the molecule. Generally, a given medium shows reciprocity that light merely behaves dextrorotation or levorotation independent of the propagation direction. While for Faraday effect, the rotation direction of polarization is relevant to whether the light propagates forward or backward the magnetic field. This non-reciprocity enables the applications in many fundamental systems, such as isolators and circulators. Similarly, the proposed CWP can induce non-reciprocal OA of Bessel beam (see Figure 2d). The rotation of the plane of polarization closely depends on the direction of propagation. From the view of the front and back sides of the CWP, the axis orientations are respectively $\mp \delta_k \rho/2$. It means that the sign of the radial frequency is reversed when changing the input side and, consequently, so is the specific rotation [$\alpha$]. To experimentally verify this non-reciprocity, the CWP is flipped to change the input direction of Bessel beam. The measured angle of polarization rotation versus propagation distance also behaves a linear variation as depicted in Figure 2e. By comparing with the case before the CWP is flipped, the rotation angle has a downward trend as propagating, and the specific rotation is measured $[\alpha]$=-1.103°/mm, showing dextrorotation.

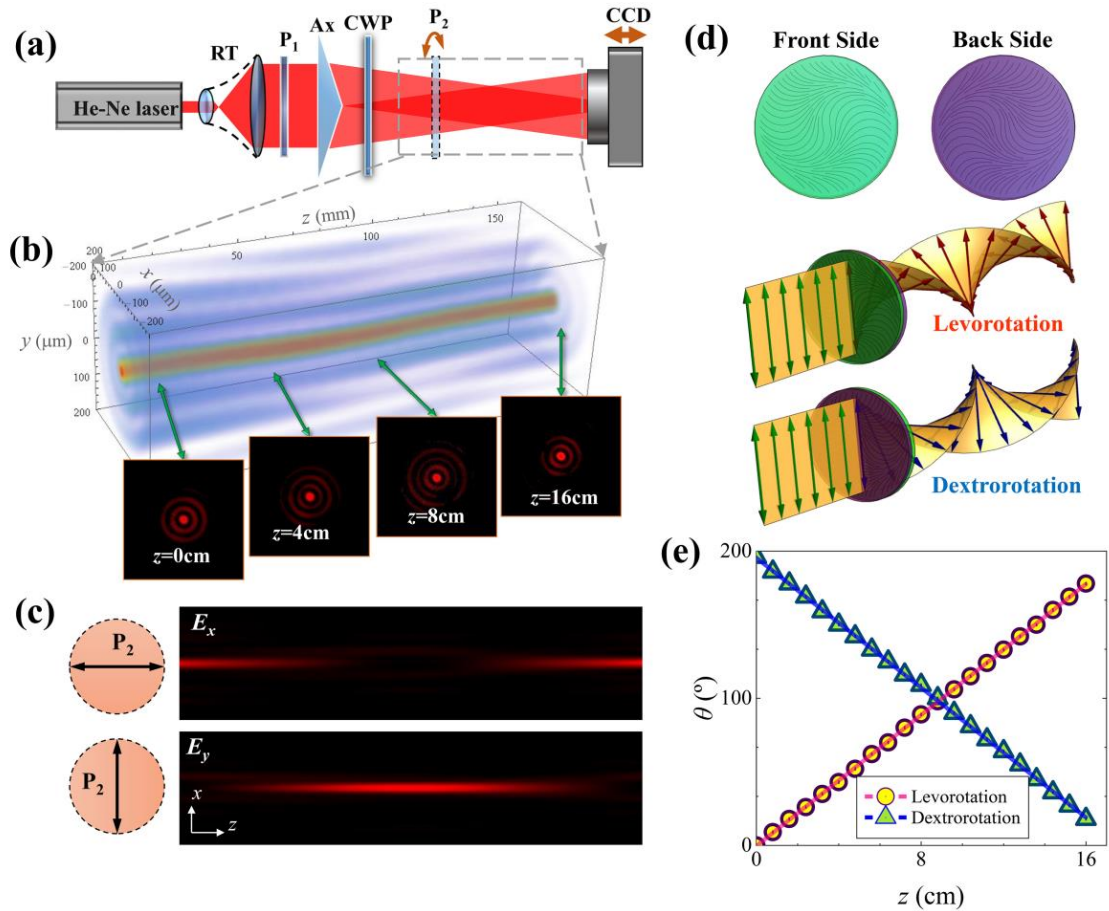

**Figure 2 Experimental demonstration of non-reciprocal OA in free space.** (a) Experimental setup for observing free-space OA. A collimated linearly polarized laser beam with a wavelength of 632.8nm passes through an axicon (Ax) and a CWP, and then is detected by an axially movable CCD camera, a polarizer ($P_1$) is used to guarantee the linearity of input polarization, an analyzer ($P_2$) is used to check the output polarization. (b) Non-diffraction propagation process captured by CCD, with the transverse intensities at different distances shown in the insets. (c) Propagation processes of (top) *x*- and (bottom) *y*-polarized components of the beam, which are obtained by inserting analyzer $P_2$ with horizontal and vertical axes, respectively. The staggered periodic extinction means that polarization rotates with propagation. (d) Schematic of the levorotation and dextrorotation by inputting from the front and back sides of CWP. (e) Measured polarization angle versus propagation distance for levorotation and dextrorotation.

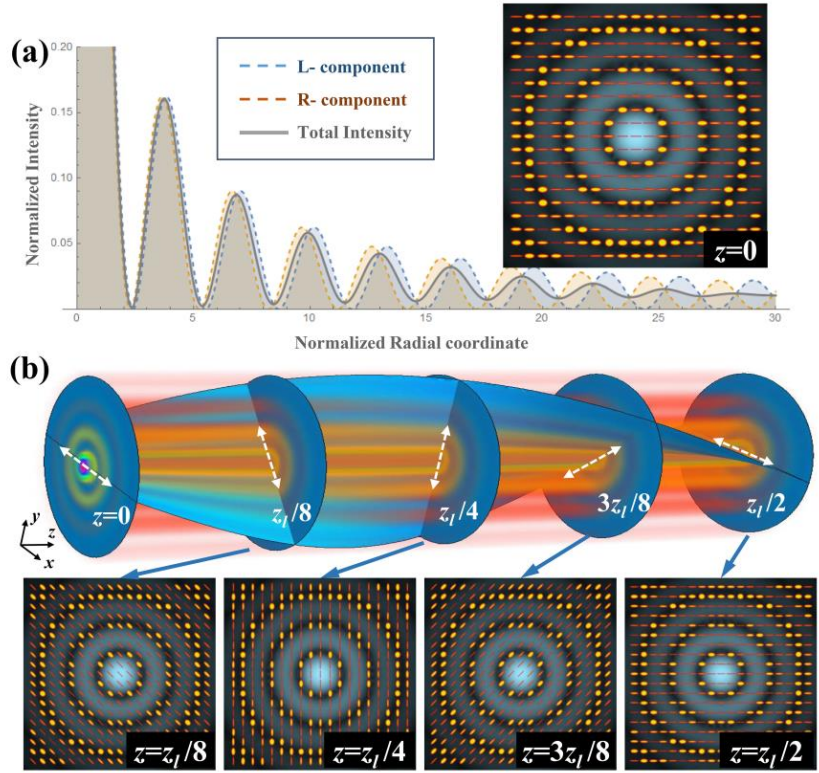

**Figure 3 Properties of optically active beam (OAB).** (a) Amplitudes profiles of the OAB and its left- and right-handed polarized components (L- and R-components). Inset shows the amplitude (background) and polarization (ellipses) distributions of the OAB. (b) Polarization rotation of the OAB during propagation within a half of rotation period $z_l$. White arrowheads denote the polarization orientation. Theoretically, an infinite-energy OAB can entirely maintain its amplitude and polarization profiles, and integrally rotates with propagation.

**Optically active beam**

The above OA can be considered as an intrinsic nature of this type of Bessel beam. Once the beam is exported from the CWP, the wave packet will never be restrained in free space, and its further evolution thereafter just depends on its internal parameters such as complex amplitude and polarization. In another word, this Beam itself has the optical rotatory power, and can be termed optically active beam (OAB). Actually, the OAB is expressed by the interference of two Bessel spin states with different radial wave vectors (see Methods, rather than Equation (1) in the approximation applied for the paraxial region), and its polarization is not exactly a uniform linear one. It has the hybrid elliptical polarization with radially varied ellipticity and constant orientation, which integrally rotates with propagation distance as shown in Figure 3. When the OAB is extinct via an analyzer, only the central energy is entirely eliminated, while a few of the surrounding one is survived. Due to the Bessel profile, the OAB performs the same propagation dynamics as that of Bessel beam, such as non-diffraction and self-healing properties.

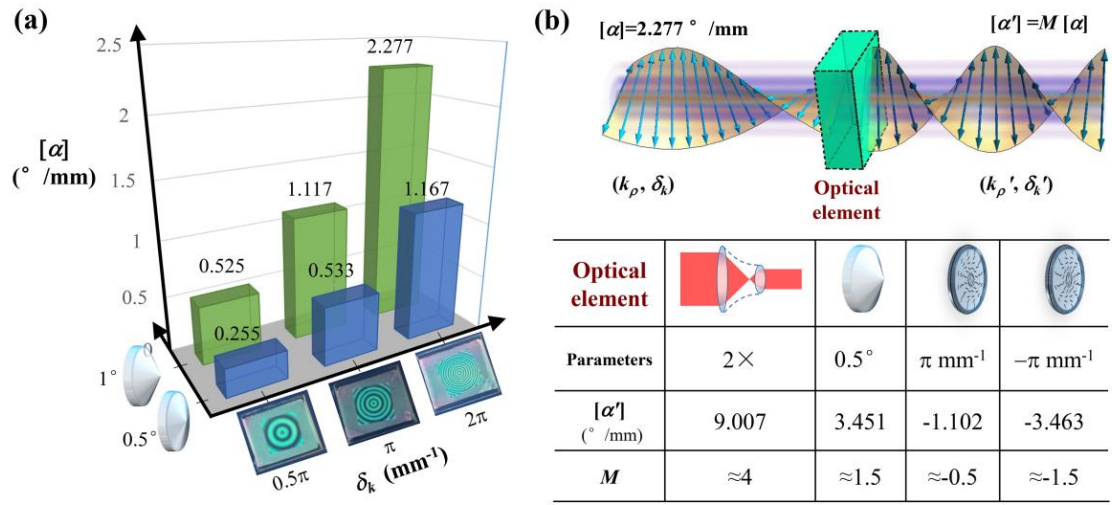

**Figure 4 Control of the rotatory power of OAB.** (a) Measured specific rotation of the OAB with different $\delta_k$ and $k_\rho$ (proportional to the physical angle of axicon). (b) Schematic of the magnification of specific rotation via an optical element (top) and the measured specific rotation and magnification $M$ (bottom table). Here the input OAB is generated by axicon of 1° and CWP with $\delta_k=2\pi$. The parameters of the optical elements in the table are magnification of beam expander (2×), physical angle of axicon (0.5°), and radial frequency $\delta_k$ of CWP ($\pi$ and $-\pi$ for the input from front and back sides), respectively. The minus $M$ means the opposite rotation of polarization.

### Control of rotatory power

There are two key parameters deciding the appearance of the OAB: the radial wave vector $k_\rho$ of input Bessel beam and the radial frequency $\delta_k$ of CWP, mainly determining the complex amplitude and polarization profiles, respectively. The optical rotatory power of the OAB (resting with the specific rotation [$\alpha$]) also depends linearly on $k_\rho$ and $\delta_k$. Here, we employ another axicon (with physical angle 0.5°) and other two CWPs ($\delta_k$=2$\pi$ and 0.5$\pi$ mm$^{-1}$) to generate OAB with different rotatory powers. Figure 4a depicts the measured specific rotation versus $k_\rho$ and $\delta_k$, and shows conformity with the expected conclusion above. Although the approximation applied for the condition $\delta_k$<<$k_\rho$<<$k$ in Equation (1) seems to restrict the upper limit of the rotatory power, the polarization rotation is always workable on the propagation axis ($\rho$=0) when $k_\rho$ and $\delta_k$ is largely increased. As a result, the specific rotation can be greatly magnified with micrographics. Interestingly, the rotatory power still can be adjusted even the OAB is already formed, by employing a special optical element, such as beam expander, axicon, and CWP (see Figure 4b), which acts as an enhancer of $k_\rho$ or $\delta_k$ in essence. Thus the magnification $M$ of [$\alpha$] equals to the product of the magnifications of $k_\rho$ and $\delta_k$ (see Methods).

**Response on refraction index**

Although the OAB is realized in air medium in the above experiments, it is also possible to exist in other non-active isotropic media, which is considered a free space, too. In a medium with refraction index $n$, the OAB only changes its wave number, and thus the specific rotation is expressed as [$\alpha_n$]=$nk_\rho\delta_k/k_0$=$n$[$\alpha$]. This means the OAB represents a faster polarization rotation in the medium, and provides a possibility to detect the refractive index of a medium. We can input an OAB into a cuvette filled with liquid medium to verify this index dependence as shown by Figure 5a. When the index is changed $\delta n$, the polarization also rotates with angle $\delta n$[$\alpha$]$L$, where $L$ is the length of the cuvette. Here, we employ the glycerol solution of various concentration from 0% to 75% as the standard liquids, and use an analyzer to check the polarization orientation of the OAB passing through the solution. Figure 5b shows the measured polarization angle versus the index of the solution, which is calibrated according to its concentration. The measured results are coincide with the theoretical prediction, and represents a linear sensibility of OA on the refractive index. Although the index resolution presented in this experiment is a bit weak (~0.0083/°), it can be majorly enhanced by magnifying the specific rotation as mentioned above.

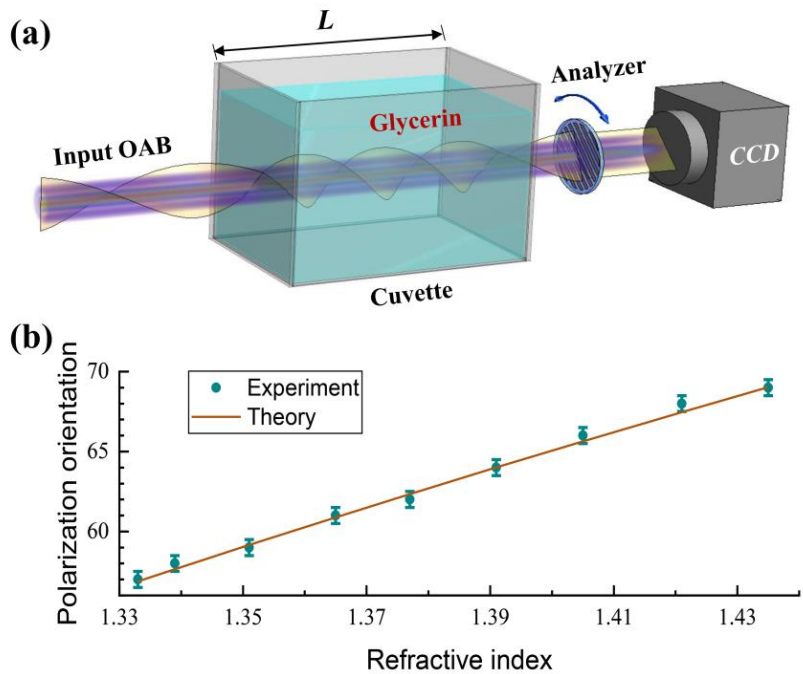

**Figure 5 Detection of refractive index change with OAB.** (a) Schematic of the setup. An OAB is input to a cuvette (L=10cm) filled with glycerol solution, passes through an analyzer, and finally captured by CCD. When changing the concentration of the solution, the polarization orientation is also changed, and can be checked by rotating the analyzer. (b) Polarization orientation versus refractive index of the solution. The OAB used here has a specific rotation $[\alpha]=1.167°$/mm.

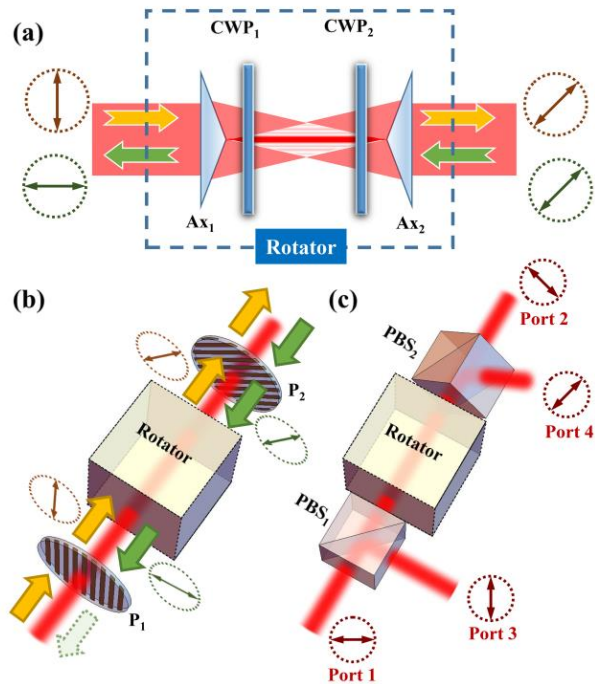

**Figure 6 Non-reciprocal optical elements based on the free-space OA.** (a) Schematic of polarization rotator consisting of a pair of axicon and CWP. The forward and backward propagations harvest polarization rotations in different directions due to the non-reciprocity of the CWP. The rotation can be adjusted by the distance between the two CWPs. Based on this non-reciprocal rotator, optical isolator (b) and four-port circulator(c) can be generated. The isolator consists of the rotator and a pair of polarizers ($P_1$ and $P_2$) at 90° and 45°, so that the dextrorotatory forward beam can pass through while the levorotatory backward beam is suppressed. The circulator consists of the rotator and a pair of polarized beam splitters, which are used to separate the polarizations in horizontal and vertical angles ($PBS_1$) and in 135° and 45° angles ($PBS_2$). Beam input in any port is only emitted from the next port in order of 1→2→3→4→1.

**Polarization rotator, isolator and circulator**

The most widely used implementation of non-reciprocal OA is the Faraday rotator, the key component of isolator and circulator. The proposed free-space OA also has the capacity to form a polarization rotator own to its non-reciprocity. The biggest issue, however, is that the polarization would be rotated without stopping as propagation because of the intrinsic OA of the beam. To unload the OA, another CWP same as that used to generate OAB has to be utilized (Figure 6a) with the purpose to slow the polarization rotation down to zero (see Methods). We experimentally verify that the OAB is reduced to a linearly polarized Bessel beam, which is clearly visualized by the dynamically analyzed image of the output field in Supplementary Movie 1. The output polarization orientation can be controlled by the distance between the two CWPs which construct a typical non-reciprocal rotator: input from different sides leads to polarization rotation in different directions. On this basis, we present a possibility to form the isolator and circulator, as illustrated by Figures 6b and 6c, respectively. For the isolator, the vertically polarized input beam obtains a polarization rotation of 45° through the rotator, while the back-propagating beam encounters an additional 45° rotation, and is isolated by the input polarizer $P_1$. Figure 6c is a typical four-port circulator to support a nonreciprocal routing of signals: light beams entering any port are only passed on to the next port in rotation order of 1→2→3→4→1. Although the proposed isolator and rotator cannot provide good isolation in the present, they theoretically offer a potential possibility for exploring novel applications of the non-reciprocal OA without media.

## Discussion

The proposed OA is also applicable to the higher-order Bessel beam or even vector Bessel beam, which has the same Gouy phase shift as the zero-order one. As a result, we can also generate the OAB with higher-order (or even vector) Bessel profile[44,45], of which the specific rotation is the same as that discussed above.

In conclusion, we demonstrated a method to possess a light beam with OA for the first time, so that polarization rotation of beam occurs in free space without requiring any active media and specially designed materials or setup. The sensibility of polarization rotation on the refractive index is further demonstrated. The OA of the beam can be selectively switched on or off by a CWP, which supports the non-reciprocal polarization rotation. This provides possibility to explore the medium-free non-reciprocal elements such as isolator and circulator.

## Methods

**Theoretical formulation of polarization rotation of Bessel beam**

Since the CWP is considered as a half wave plate with space-variant fast axes $-\delta_k\rho/2$, its transmission matrix is expressed as

$$\mathbf{T}(\rho) = \begin{bmatrix} \cos\delta_k\rho & -\sin\delta_k\rho \\ -\sin\delta_k\rho & -\cos\delta_k\rho \end{bmatrix}. \tag{2}$$

The CWP has different responses for the two orthogonal circular polarizations. For an incident wave with circular polarization, the corresponding output fields are

$$\begin{cases} \left|\mathbf{E}_{\text{out}}^{\sigma_+}\right\rangle = \mathbf{T}\left|\boldsymbol{\sigma}_+\right\rangle = e^{i\delta_k\rho}\left|\boldsymbol{\sigma}_-\right\rangle \\ \left|\mathbf{E}_{\text{out}}^{\sigma_-}\right\rangle = \mathbf{T}\left|\boldsymbol{\sigma}_-\right\rangle = e^{-i\delta_k\rho}\left|\boldsymbol{\sigma}_+\right\rangle \end{cases}, \tag{3}$$

where $|\sigma_-\rangle$ and $|\sigma_+\rangle$ are the Jones vectors of circular polarizations (spin states), expressed as $[1,\pm i]^T/\sqrt{2}$. It can be evidently seen that the right and left spins are respectively transformed into left and right ones, with attached different Pancharatnam-Berry phases $\exp(\pm i\delta_k\rho)$.

For a monochrome input zero-order Bessel beam with linear polarization propagating along z-axis, the complex amplitude is described by $\mathbf{E}_0=J_0(k_\rho\rho)|\mathbf{L}(\theta_0)\rangle$, where $J_0$ is the zero-order Bessel function of the first kind, $k_\rho$ is a real constant describing the radial component of wave vector, $|\mathbf{L}(\theta_0)\rangle=[\cos\theta_0, \sin\theta_0]^T$ denotes the linear polarization along angle $\theta_0$. The Gouy phase shift of the Bessel beam is expressed by $\varphi_{Gouy}(k_\rho)=k_\rho^2 z/(k_0+\sqrt{k_0^2-k_\rho^2})$, where $k_0=2\pi/\lambda$, $\lambda$ is the wavelength in the free space[38]. It means that the Bessel beam shows a linearly propagation-varied phase distinct with the dynamic phase, closely dependent on the radial wave vector.

When the Bessel beam launches to the CWP, the output field is written as $|\mathbf{E}_{out}\rangle=\mathbf{T}\cdot\mathbf{E}_0=J_0(k_\rho\rho)[\exp(-i\theta_0-i\delta_k\rho)|\sigma_+\rangle+\exp(i\theta_0+i\delta_k\rho)|\sigma_-\rangle]/\sqrt{2}$. Noting that the Bessel beam can be paraxially approximated by axicon phase function, i.e., $J_0(k_\rho\rho)\approx\exp(-ik_\rho\rho)$, the output field can be written as $|\mathbf{E}_{out}\rangle=J_0(k_+\rho)\exp(-i\theta_0)|\sigma_+\rangle/\sqrt{2}+J_0(k_-\rho)\exp(i\theta_0)|\sigma_-\rangle/\sqrt{2}$, where $k_\pm=k_\rho\pm\delta_k$. Taking into consideration the Gouy phase shift, we get

$$\left|\mathbf{E}_{out}\right\rangle = J_0(k_+\rho)|\sigma_+\rangle e^{i\varphi_{Gouy}(k_+)-i\theta_0}/\sqrt{2}+J_0(k_-\rho)|\sigma_-\rangle e^{i\varphi_{Gouy}(k_-)+i\theta_0}/\sqrt{2}. \tag{4}$$

The right- and left-handed polarization components of the field passing through the CWP acquires a phase difference

$$\Delta\varphi = \varphi_{Gouy}(k_+)-\varphi_{Gouy}(k_-)-2\theta_0 = \frac{2k_\rho\delta_k}{k'}z-2\theta_0, \tag{5}$$

where $k'=(\sqrt{k_0^2-k_+^2}+\sqrt{k_0^2-k_-^2})/2$. Under an actual condition of $\delta_k\ll k\ll k_0$, the approximation $k'\approx k_0$ can be deduced. This indicates that a birefringence-like effect occurs for the two circular polarizations.

Ignoring the common phase terms, the output field is the coherent superposition of two Bessel beams with different radial wave vectors and circular polarizations

$$\left|\mathbf{E}_{out}\right\rangle = J_0(k_+\rho)e^{i\Delta\varphi/2}|\sigma_+\rangle + J_0(k_-\rho)e^{-i\Delta\varphi/2}|\sigma_-\rangle \tag{6}$$

In the paraxial region, if $\delta_k$ is small enough, the approximation $J_0(k_+\rho)\approx J_0(k_-\rho)\approx J_0(k_\rho\rho)$ (see Figure 4a where $\delta_k=0.1k_\rho$) can be applied, then the output field is given by Equation (1).

**Fabrication of conical-wave plate**

To experimentally demonstrate the OA of Bessel beam, we fabricate the proposed CWP in a typical birefringent material LC. A photoalignment technique based on sulfonic azo-dye SD1 and dynamic micro-lithography[39,40] is employed to obtain the LC CWP. As shown in Figure 7a, the LC cell is composed of two pieces of SD1-coated indium-tin-oxide (ITO) glass substrates and nematic LC E7. The photoalignment agent SD1 tends to reorient its absorption oscillators perpendicular to the polarization direction of the illuminated UV light, and then guides the LC director by intermolecular interaction[41]. Via a multi-step partly-overlapping exposure process[42], the conically-distributed alignment of LC is realized. The tilting of LC can be controlled by applying proper voltage, thus the effective refractive index and phase retardation are tunable. Here, ~2.5 volts square-wave voltage is applied on the LC to achieve the half-wave condition

Figures 7b shows the micrograph of the LC CWP (about 8 mm in square) with the parameter $\delta_k = \pi$ mm$^{-1}$ under a polarized optical microscope. Due to the birefringence of LC, the optical axis orientation of CWP can be manifested by a pair of crossed polarizers, as shown as the concentric-ring-like fringes. The cyanic color of the micrograph is attributed to the corresponding wavelength meeting the half-wave condition. The bright areas correspond to the axes along ~45° or 135° with respect to the polarizer of the microscope, while dark areas correspond to ~0° or 90°. The adjacent bright fringes reflects the axis variation of 90°, and the fringe spacing is ~1 mm, almost matching the designed axis orientation $\Theta = -\delta_k \rho / 2$. To quantitatively characterize the CWP, the Pancharatnam–Berry phase method[43] is adopted to measure the Stokes parameters of a linearly polarized beam passing through the CWP, and the distribution of the polarization angle which equals $-\delta_k \rho$ is calculated as shown in Figure 7c. The top inset depicts the distribution curve on the dashed transversal line, revealing the conical arrangement of the optical axis. By utilizing experimental data fitting, the coefficient $\delta_k$ is measured as $\delta_k$ =0.9541$\pi$ mm$^{-1}$, which matches well with our theoretical design.

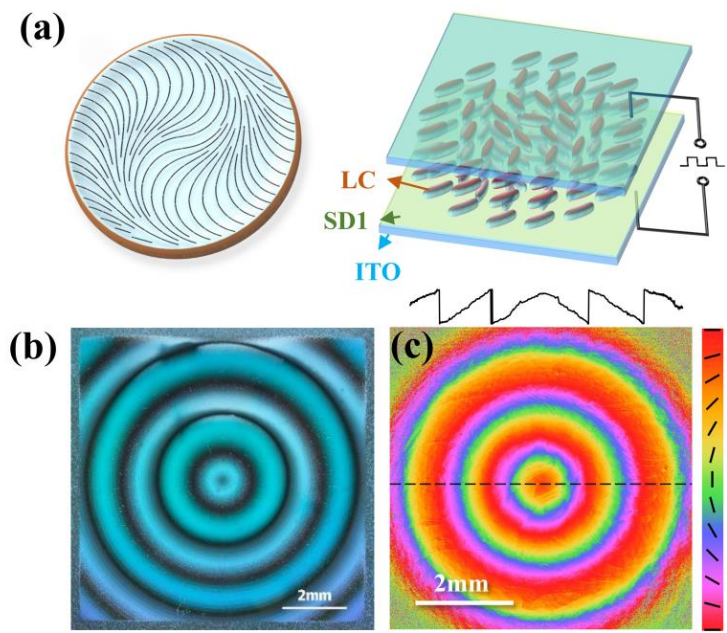

**Figure 7 Theoretical and measured structure of CWP.** (a) Schematic of the fast axis distribution (left) and the liquid crystal cell configuration (right) of CWP; (b) Micrograph of the CWP; (c) Measurement result of polarization angle of a linearly polarized beam passing through the CWP, where the distribution on the intersecting line is depicted by the inserted curve.

**Magnification of specific rotation**

The specific rotation can be magnified by optical elements including beam expander, axicon, and CWP, etc. This magnification is essentially imposed to the radial wave vector of Bessel beam and the radial frequency of the CWP. Supposing the specific rotation of an OAB is expressed as [$\alpha(k_\rho,\delta_k)$], and after through the optical elements, a transforming of parameters happens from ($k_\rho,\delta_k$) to ($k_\rho',\delta_k'$), the specific rotation becomes [$\alpha'$]=[$\alpha(k_\rho',\delta_k')$]= $k_\rho'\delta_k'$[$\alpha$]/$k_\rho\delta_k$, namely, the magnification is $M= k_\rho'\delta_k'/k_\rho\delta_k$. For example, by using a reversed beam expander with magnification 2×, the OAB is reduced by half both for the beam size and the polarization distribution, i.e. $k_\rho'=2k_\rho$, and $\delta_k'=2\delta_k$. Thus the magnification $M=4$. By using an axicon, the radial wave vector is changed to $k_\rho'=k_\rho+k_{\rho 1}$, where $k_{\rho 1}$ is relevant to the physical angle of axicon, and thus $M=1+k_{\rho 1}/k_\rho$. When the OAB passes through a CWP with fast axis orientation $-\delta_{k1}\rho/2$, the spin states are converted to each other, and the output field is expressed as

$$|\mathbf{E}_{out}\rangle = J_0(k'_-\rho)e^{i\Delta\varphi'/2}|\sigma_+\rangle + J_0(k'_+\rho)e^{-i\Delta\varphi'/2}|\sigma_-\rangle, \qquad (7)$$

where $k'_\pm=k_\rho\pm(\delta_k-\delta_{k1})$, and $\Delta\varphi'=\varphi_{Gouy}(k'_-)-\varphi_{Gouy}(k'_+)$. Thus the specific rotation can be deduced as [$\alpha'$]=[$\alpha(k_\rho,\delta_{k1}-\delta_k)$], and the magnification $M=\delta_{k1}/\delta_k-1$. Here, a minus magnification means that the rotating direction is reversed. It is important to note that when $\delta_{k1}=\delta_k$, the magnification is zero, showing that no polarization rotation happens. Namely, the OA of the beam can be eliminated by employing a same CWP as that used to generate the OAB.

**ACKNOWLEDGEMENTS**

Thanks Dr. Biqiang Jiang of Northwestern Polytechnical University providing the sample of glycerol solution. This work was supported by the National Key R&D Program of China (2017YFA0303800); Joint Fund of the National Natural Science Foundation Committee of China Academy of Engineering Physics (NSAF) (U1630125); and National Natural Science Foundations of China (NSFC) (11634010, 61675168, 11774289, 91850118，and 11804277).